\documentstyle[prl,aps,multicol,epsf]{revtex}

\begin{document}

\bibliographystyle{simpl1}

\title{Intrinsic Decoherence in Mesoscopic Systems}

\author{Maxim Vavilov and Vinay Ambegaokar} 

\address{Laboratory of Atomic and Solid State Physics,
Cornell University, Ithaca NY 14853, USA}
\maketitle

\centerline{(September 20, 1997)}

\begin{abstract}
We point out that even at the absolute zero of temperature  
environmental decoherence limits the destructive
interference between time-reversed paths 
for an electron in a disordered metal, and thus causes 
the leading (`weak localization') quantum correction to the conductivity
to saturate at $T=0$.  Our
calculation, which is intended to be illustrative rather than  
complete, uses a model in which an electron interacts with the
fluctuations of the mean voltage in the sample.  The 
average of the fluctuations produces the steady damping well known in
Brownian motion, introduces a direction of time, ensures that
arbitrarily long time-reversed electron paths lose phase coherence, and
is consistent with the experimental observation of a finite low
temperature phase-breaking lifetime. 
\end{abstract}
\draft


\begin{multicols}{2}
A recent experiment by Mohanty, Jariwala, and Webb \cite{1}
has called into question conventional wisdom about the first quantum
correction to the electrical conductivity of disordered metals.
This correction has been studied for 20 years and named weak localization
\cite{2}.  The physical origin of the correction is the coherent
back-scattering of electrons due to the quantum interference of
time-reversed paths.  The theory contains  
a phase-breaking lifetime which limits the length of
destructively interfering paths.  Calculations to date have this lifetime
tending to infinity as the temperature approaches zero, but the
experiments indicate that it remains finite.

In our opinion the previous theoretical work has not included the
full effect of
environmental fluctuations.  We point out that the environment has
an average dissipative effect, well known from the theory of Brownian
motion, which survives at $T=0$ and appears to resolve the controversy.  
Our aim is to elucidate the missing physics using a simple model, leaving
more detailed questions for further study.   

We suppose that the electrons in the sample move in the electric field
produced by fluctuations of the voltage in the electric circuit which
contains the sample, and we consider the destruction of electron 
self-interference due to Nyquist noise at temperature $T$.  We use the
method of Feynman
and Vernon \cite{3} as elaborated by Leggett and Caldeira \cite{4}, and
describe the electric field by a set of
oscillators whose density is determined from the
fluctuation--dissipation theorem.   Averaging over the environmental
oscillators yields the Feynman-Vernon influence functional which is the
exponential of an action with both real and imaginary parts.  For the
purposes at hand, the real part plays the essential role; the
contribution of the imaginary part---which in the classical limit is the
effect of a fluctuating Langevin force \cite{5}---to phase-breaking does
indeed vanish at zero temperature.  

The weak localization correction to the conductivity is, in a standard
approach, proportional to an integral over time $t$ of the
probability $P(t)$ that electron starting from some point at $t=0$ returns
to the same point at time $t$.  We use this starting point, as
developed by Chakravarty and Schmid \cite{7}, introducing the random
impurity potential averaged over microscopic length scales and then
integrating over the different quasi-classical trajectories. 

At low temperatures the decoherence time due to Nyquist noise 
is the slowest function of temperature, and only one
source of the destruction of the coherence. 
Besides, we limit ourselves here to the case of quasi one 
dimensional systems.  Our result (see Eq.(\ref{15}))
appears to
apply to some of the data presented in \cite{1}, but
it is worth mentioning again that we consider this to be a model
calculation, designed to illustrate some essential physics.  Important
other sources
of decoherence such the electron-electron and the electron-phonon
\cite{8,9} interaction have been omitted. 

We consider a system described by the Hamiltonian
\begin{eqnarray}
\label{1}
H=H_e+H_R+H_{e-R},
\end{eqnarray}
where
\begin{eqnarray}
\label{2}
H_e=\frac{p^2}{2M}
\end{eqnarray}
is the Hamiltonian of the free electron, and
\begin{eqnarray}
\label{3}
H_R=\frac{1}{2}\sum_{\alpha}\left(\frac{p_{\alpha}^2}{m_{\alpha}}+
m_{\alpha}\omega_{\alpha}x^2_{\alpha}\right)
\end{eqnarray}
is the Hamiltonian of the environment, assumed to consist 
oscillators with a distribution to be given later. The
interaction
of the environment and the electron has the form
\begin{eqnarray}
\label{4}
H_{e-R}=-q\frac{e}{L}\sum_{\alpha}x_{\alpha}.
\end{eqnarray}
The environmental degrees of freedom are the harmonics of the 
voltage across the sample.  In Eq. (\ref{4}) $L$ is the length of
the sample and $q$ denotes the coordinate of the electron.

Because we are interested in the motion of the electron only we should
average the density matrix over the environmental degrees of freedom. 
Then we
get \cite{3,4,5,10}
\begin{eqnarray}
\label{5}
\begin{array}{c}
P(t) =\rho(q_0,q_0,t)=\\ [12pt]\displaystyle
\int_{q_0}^{q_0} {\cal D} q(\tau)\int_{q_0}^{q_0} {\cal D} q'(\tau)
\exp\Big(i(S_0[q]-S_0[q'])\Big)F[q,q'],
\end{array}
\end{eqnarray}
where $S_0$ is the action for a free particle, 
\begin{eqnarray}
\label{6}
F[q,q']=\exp\big(iS_1[q,q'] - S_2[q,q']\big)
\end{eqnarray}
is the influence functional, and $S_1[q,q']$ and $S_2[q,q']$ are the real 
and imaginary parts of the effective action,
\begin{eqnarray}
\label{7}
\begin{array}{c}
\displaystyle
S_1[q,q']=\int_0^tds\int_0^s du \sum_{\alpha}\frac{e^2}
{2m_{\alpha}\omega_{\alpha}L^2}
\times\\[12pt]\displaystyle
\sin(\omega_{\alpha}(s-u))
\left(q(s)-q'(s)\right)
\left(q(u)+q'(u)\right),
\end{array}
\end{eqnarray}
\begin{eqnarray}
\label{8}
\begin{array}{c}
\displaystyle
S_2[q,q']=\int_0^tds\int_0^s du \sum_{\alpha}\frac{e^2}
{2m_{\alpha}\omega_{\alpha}L^2}\coth\frac{\omega_{\alpha}}{2T}\times
\\[12pt]\displaystyle
\cos(\omega_{\alpha}(s-u))\left(q(s)-q'(s)\right)\left(q(u)-q'(u)\right)
.
\end{array}
\end{eqnarray}

The imaginary part $S_2$ of the effective action 
contains the correlator of the
environment field, 
quantum or classical, but in the classical case we have instead of the
function
$\coth(\omega/2T)$,
its approximation at small values of the argument,
i.e.
$2T/\omega$. This approximation is appropriate when the temperature of
the
system is larger than the characteristic frequencies of the system.
At lower temperature, the more accurate approximation is
\begin{equation}
\label{8.5}
\coth(\omega/2T)\approx{\rm sign}\omega+2T/\omega. 
\end{equation}
 One can view the
first term in this equation as coming from the zero-point motion of the
oscillators.  We shall see below that it makes no contribution to the
return probability, thus calling into question the attribution in ref.
\cite{1} of the observed saturation to `zero point fluctuations.'
By contrast, the real part of the effective action does not depend 
on the temperature of the system and, as we shall see, it
contributes to the zero temperature behavior of the system. This factor
describes the average damping effect of the fluctuating field.

The summation over the environmental modes $\alpha$ may be replaced
by an integration over frequency of the density of modes
of the voltage, defined by
$J(\omega)=2\pi \sum_{\alpha}\delta(\omega-
\omega_{\alpha})/2m_{\alpha}\omega_{\alpha}$.  This can in turn be determined 
from the fluctuation--dissipation theorem. The correlator of the voltage 
is $<x(t)x(0)+x(0)x(t)>/2$, where $x(t)=\sum_{\alpha}x_{\alpha}(t)$;
the 
average, with respect to the Gibbs distribution corresponding to
Eq. (\ref{3}), can be calculated explicitly.  From the 
fluctuation dissipation theorem, the 
correlator is proportional to the susceptibility of the field, here
given by the impedance of the electric circuit which we take for
simplicity to be purely resistive. As a result 
\begin{eqnarray}
\label{9}
J(\omega)=\omega R_{eff}. 
\end{eqnarray}

We can now verify that the term ${\rm sign}\omega$ in Eq. (\ref{8.5})
does not contribute
to the motion by performing an integration over $\omega$ in Eq. (\ref{8}) from
$-\omega_c$ to $\omega_c$, where $\omega_c$ is the cutoff \cite{footnote}
.
Then we get a sharply peaked function of $(s-u)$ whose integral over
the difference $(s-u)$ is exactly zero. [In performing this integral we
may neglect the deviations
of $q(s)$ and $q'(u)$, since they may be considered slow functions of the
difference $(s-u)$ on the scale $(\omega_c)^{-1}$.] After these manipulations
the imaginary part of the action is
\begin{eqnarray}
\label{10}
S_2[q,q']=\frac{e^2 T}{L^2}R_{eff}\int_0^t ds \left(q(s)-q'(s)\right)^2.
\end{eqnarray}
With the choice (\ref{9}) for $J(\omega)$ the integral over modes can also be
done for the real part of the action, yielding
\begin{eqnarray}
\label{11}
\begin{array}{c}
\displaystyle
S_1[q,q']=-\frac{e^2}{2L^2}R_{eff}\int_0^t ds \left(q(s)+q'(s)\right)\times
\\[12pt]
\displaystyle
\frac{d}{ds}\left(q(s)-q'(s)\right),
\end{array}
\end{eqnarray}
where we have ignored a singular term proportional to $\int_0^t ds
\left(q^2(s)-q^{\prime 2}(s)\right)$ which makes no contribution to the time
reversed paths considered below.\cite{ft2}

Following Chakravarty and Schmid \cite{7}, we calculate the
propagation of the electron in the field of random impurities, now,
however,  with the addition of the environmental interaction. 
We assume that there is a hierarchy
of length scales $\lambda_F\ll l_0 \ll l(T)$, where $\lambda_F$ is the
electron wavelength, $l_0$ is the impurity mean free path, and $l(T)$ is
the environmental decoherence length.  We suppose that the motion
between impurity collisions is described by the effective action
given by Eqs. (\ref{5}-\ref{11}).  Since this action is quadratic in the
electron co-ordinate, it can in principle be done exactly \cite{5},
leaving a double sum over impurity positions.  We argue as in ref.
\cite{7} that because of rapidly oscillating exponentials only  
time reversed paths contribute to this double sum, and then replace the
single sum over trajectories by an average weighted by the probability of
diffusive motion in the random impurity potential.  It is now convenient
to set the zero of time
in such a way that the integration is over a symmetric interval.
Then the 
relation between $q(\tau)$ and $q'(\tau)$ is $q'(\tau)=q(-\tau)$,
with the weight of each trajectory given by $\exp(-\int_{-t/2}^{t/2}dt
\dot{q}^2(\tau)/4D)$, where $D$ is the diffusion constant. It is 
convenient to introduce new variables 
$x(\tau)=(q(\tau)+q(-\tau))/\sqrt{2}$ and
$\xi(\tau)=(q(\tau)-q(-\tau))/ \sqrt{2}$. Then the return probability is
given by the expression
\begin{eqnarray}
\label{12}
\begin{array}{c}
\displaystyle
P(t)=\int_{-\infty}^{\infty}\frac{dx_0}{\sqrt 2}\int_{x_0}^{x_f} {\cal D}
x(\tau) 
\int_0^0 {\cal D}\xi(\tau)\times \\[12pt]\displaystyle
e^{-\frac{1}{4D}\int_0^{t/2}\dot{x}^2(\tau)}
e^{-\frac{1}{4D}\int_0^{t/2}\dot{\xi}^2(\tau)}\times \\[12pt]\displaystyle
\exp{\left(-4R_{eff}T\frac{e^2}{L^2}\int_0^{t/2}dt \xi^2(\tau)\right)}
\times \\[12pt]\displaystyle
\exp{\left(-2iR_{eff}\frac{e^2}{L^2}\int_0^{t/2}dt \dot{\xi}(\tau)
x(\tau)\right)}
.
\end{array}
\end{eqnarray}
The integrals in this expression may be done successively.  First we do the
functional integral over $x(\tau)$.  We find that the end points $x_0$ and
$x_f$ then only occur in a Gaussian function of $(x_0 - x_f - K)$ so that
$K$, a constant depending on the path $\xi (\tau)$, and $x_f$  drop out after a
trivial
integration over $x_0$.  Finally one is left with a functional integral
over $\xi(\tau)$  equivalent to that for a harmonic oscillator.  The final
answer is
\begin{eqnarray}
\label{13}
P(t) = \sqrt{\frac{\Omega}{8\pi D \sinh(\Omega t/2)}},
\end{eqnarray}
with
\begin{eqnarray}
\label{13.5}
 \Omega^2 = 16\left(\frac{D}{L^2}\frac{e^2 R_{eff}}{\hbar}\right)
\left[\frac{D}{L^2}\frac{e^2 R_{eff}}{\hbar} + \frac{T}{ \hbar}\right].
\end{eqnarray}
In the last equation the first (temperature independent) term comes from
the real part of the action whereas the second comes from the imaginary
part and vanishes at zero temperature ($T\rightarrow 0$). At zero
temperature, the driven response of the environment provides the 
friction \cite{5} which breaks the symmetry with respect to time
reflection and gives different contributions to the action of a
trajectory and its time reflection. 

The correction to the conductivity is the integral with respect to time of
the return probability,
\begin{eqnarray}
\label{14}
\Delta \sigma=-\frac{2e^2}{\pi\hbar}D\int_0^{\infty}dt P(t).
\end{eqnarray}
This integral converges at both limits at any temperature. Making the 
integral dimensionless we get 
\begin{eqnarray}
\label{15}
\Delta \sigma=-\alpha\frac{e^2}{\pi\hbar}\sqrt{\frac{D\hbar}{T_q}}
\left({\frac{T_q}{T+T_q}}\right)^{1\over 4},
\end{eqnarray}
where 
\begin{eqnarray}
\label{16}
\alpha=\frac{1}{\sqrt{2\pi}}\int_0^{\infty}\frac{d\xi}{\sqrt{\sinh \xi}}
\approx 1.5
\end{eqnarray}
and 
\begin{eqnarray}
\label{17}
T_q=\frac{D}{L^2}e^2 R_{eff}.
\end{eqnarray}
We see that there exists a characteristic temperature $T_q$ below which 
the behavior of the system no longer depends on the environment temperature.
In our model this temperature is proportional to the inverse of the time 
it takes an electron to diffuse through the sample
divided by the number of conducting channels. 
At high temperatures $T\gg T_q$, the correction to the conductivity
Eq. (\ref{15}) is in agreement with the result of ref. \cite{7}, and was
observed in one dimensional samples in ref. \cite{1}.

To make connection with the usual expression for the correction
to the conductivity in one dimension, we may define a quantity
$\tau(T)$ so that Eq. (\ref{15}) takes the form
 \begin{eqnarray}
\label{18}
\Delta\sigma(T)= - \frac{e^2}{\pi\hbar}\sqrt{D\tau(T)}.
\end{eqnarray}
in the present model $\tau(T)$, which is usually referred to as the
decoherence time of the electron system,	
is of the order $\Omega^{-1}$, with $\Omega$ given by 
(\ref{13.5}). 

The expression for the $\tau(T)$ has a universal form in the sense that
it is completely described by one free parameter. If we choose this
parameter to be the decoherence time at zero temperature, then
\begin{eqnarray}
\label{19}
\tau(T)=\tau_0\sqrt{\frac{1}{1+T\tau_0/\alpha^2\hbar}}.
\end{eqnarray}
This behavior of $\tau(T)$
is in qualitative accord with the experimental observations reported in 
\cite{1} and references therein.

In summary, we have shown in a simple model that there is a temperature
independent
environmental decoherence effect which limits the backscattering of
electrons in a disordered metal.  We suggest that this is
the essential physics behind the observed saturation of the weak
localization correction to the conductivity.  

\begin{acknowledgements}
This work is supported in part by the NSF under grant DMR-9407245.
\end{acknowledgements}

\end{multicols}


\begin{references}
\bibitem{footnote} Our problem has
the  natural cutoff $\omega_c\sim 1/\tau_0$, where $\tau_0$
is the elastic mean free time.
\bibitem{ft2}  In the Leggett-Caldeira theory this singular piece is
canceled by a counter term.  Here, its presence or absence does not
affect the physics
\end{references}
\end{document}